\documentclass{mem}
\usepackage{natbib}\usepackage{txfonts}\usepackage{balance}
\usepackage{graphicx}
\usepackage{txfonts}
\usepackage[a4paper]{hyperref}
\idline{79}{3}
\begin{document}
\def\teff{$T\rm_{eff }$}
\def\kms{$\mathrm {km s}^{-1}$}

\title{
NGC 2419:\\ an ``intergalactic wanderer'' or a simple Galactic Globular Cluster? 
}

   \subtitle{}

\author{ 
M. \,Di Criscienzo$^1$,
C. \, Greco$^3$,
M.\,Dall'Ora$^2$,
V.\, Ripepi$^2$,
G.\, Clementini$^3$,
M.\, Marconi$^2$,
L.\, Federici$^3$,
L.\, Di Fabrizio$^4$,
I.\, Musella$^2$,
L.\, Baldacci$^3$,
M.\, Maio$^3$
          }
\offprints{M. Di Criscienzo}
\institute{
$^1$INAF- Osservatorio Astronomico di Roma, Monteporzio Catone,(RM) Italy\\
 \email: dicrisci@na.astro.it\\
$^2$INAF-Osservatorio Astronomico di Capodimonte, Napoli, Italy\\
$^3$INAF-Osservatorio Astronomico di Bologna, Italy\\
$^4$INAF- Telescopio Nazionale Galileo, Santa Cruz de La Palma, Spain}
 
\authorrunning{M. Di Criscienzo}

\titlerunning{NGC 2419}

\abstract{
We have carried out a new photometric study of the remote Galactic globular cluster NGC 2419, using proprietary and archive $B,V,I$ time-series CCD photometry of the cluster, that allowed us to discover a large number of new variable stars and to obtain a new colour magnitude diagram that reaches $V\sim$ 26 mag over a field of 50$x$43 square arcmin centered on NGC 2419. The new 
variables include 39 RR Lyrae and 11 SX Phoenicis stars.
The pulsation properties of the new RR Lyrae stars confirm and strengthen the classification of NGC 2419 as an Oosterhoff type II cluster.  
}
\maketitle{}

\section{Introduction}

\begin{table*}
\fontsize{8}{8}
\caption{Instrumental set-ups and logs of the observations}
         \label{t:obs}
$$         \begin{array}{llllllrrr}
           \hline
{\rm Dates}                & {\rm Telescope}     & {\rm ~~Instrument} & {\rm Detector}  &{\rm ~Resolution}  & {\rm ~~~~FOV}     &{\rm N_ B} &{\rm N_ V}& {\rm N_ I}\\
                           &                     &                    &                 &                   &                   &           &          &           \\
{\rm UT}                   &                     &                    &{\rm (pixel})    &{\rm ~~(\prime \prime/pixel)}&         &           &           &         \\
            \noalign{\smallskip}
            \hline
            \noalign{\smallskip}
{\rm Sep.,2003 - Feb., 2004} & {\rm TNG~~~} & {\rm  ~~~Dolores }   &{\rm 2048\times2048}     &~~~0.275&9.4^{\prime}\times\ 9.4^{\prime}& 20~&22~&-\\
{\rm May, 1994 - Mar., 2000} & {\rm HST~~~} & {\rm  ~~~WFPC2   }   &{\rm 3\times800\times800}&~~~0.1&3\times2.5^{\prime}\times\ 2.5^{\prime}& -~&18~&10\\
{\rm Nov., 1997}             & {\rm HST~~~} & {\rm  ~~~WFPC2   }   &{\rm 3\times800\times800}&~~~0.1&3\times2.5^{\prime}\times\ 2.5^{\prime}& -~&7~~&39\\
{\rm Dec., 2002}             & {\rm SUBARU}& {\rm  Suprime-Cam}&{\rm 2048\times4096}     &~~~0.20&~34^{\prime}\times\ 27^{\prime} & -~ &165~&16\\
\hline
\end{array}
$$
\end{table*}

NGC 2419 is one of the most distant and luminous globular clusters (GCs) in the Milky Way 
(MW, $R_{GC}$=90 kpc,   Harris et al. ,1997) but has several unusual properties for an 
outer halo GC; in particular with  M$_V$ = $-$ 9.5 mag (Harris, 1996) and [Fe/H]= -2.1 dex  
it is much more luminous and metal-poor than the majority of the other outer halo clusters. 
The cluster horizontal branch (HB) also resembles that of much closer ``canonical'' metal-poor 
clusters like M15 or M68, and previous investigations on this cluster show that NGC 2419 has 
the same age of M92, within 1 Gyr (Harris et al., 1997).  However, NGC 2419 is not an inner 
halo cluster migrated out on an elliptical orbit, since its dynamical parameters and orbital 
properties are typical of an outer halo cluster. NGC 2419 is also anomalous  in the 
half-light radius (Rh) vs M$_V$ plane, where occupies the same 
strange position of M54 and $\omega$ Cen. In fact they all look significantly looser 
that expected for their brightness.
All these peculiarities and the similarity with $\omega$ Cen and M 54  suggest 
that NGC 2419 could have an extragalactic origin and be the relict of a dwarf 
galaxy tidally disrupted by the MW (Mackey $\&$ Van den Berg, 2004); from 
here the appellative ``Intergalactic Wanderer'' used in the literature. 
Our project aims to study in detail, both the cluster color-magnitude diagram (CMD) 
and the variable star population, using $B,V$ and $I$ time-series CCD photometry covering 
an area that extends well beyond the cluster published tidal radius, in order to 
verify whether multiple stellar populations and tidal tails exist in the cluster,
and to check whether the properties of the RR Lyrae stars support an extragalactic 
origin for NGC2419.
\begin{figure}[t!]
\resizebox{\hsize}{!}{\includegraphics[clip=true]{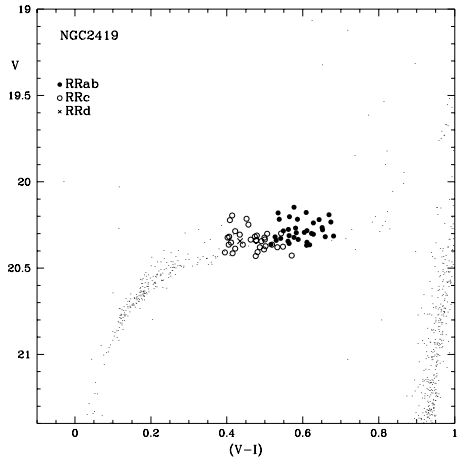}}
\caption{\footnotesize
Portion of the cluster $V$ vs $V-I$ CMD, from the SUBARU dataset, zoomed at the level 
of the HB, with the RR Lyrae stars overplotted using filled and open symbols for 
fundamental-mode and first-overtone pulsators, respectively.}
\label{eta}
\end{figure}
\section{Observation and data reduction }
Logs of the observations used for the present investigation  and details of the instrumental set-up at the various telescopes are provided in Table \ref{t:obs}.
Images were pre-reduced following standard techniques (bias
subtraction and flat-field correction) with IRAF.
We measured the star magnitudes by PSF-fitting photometry, running  
 the DAOPHOTII/ALLSTAR/ALLFRAME packages \citep{stetson87,stetson94} on 
 the TNG, HST and SUBARU datasets, separately. 
 Typical internal errors of the $V$ band photometry for
single phase points at the level of the HB are in the range from 0.01 to
0.02 mag.\\
The large field of view of the Suprime-Cam (34 $\times$ 27 arcmin$^2$) and   
dithering of the telescope pointings resulted in the survey of a total area 
of 50 $\times$ 43 arcmin$^2$ centered on NGC~2419, 
including both the TNG and HST fields.
The absolute photometric calibration was obtained by using local
standards in NGC~2419 from P.B.
Stetson's list\footnote{Available at http://cadcwww.dao.nrc.ca/standards/}.
Fig.1 shows the $V, V-I$ CMD of NGC 2419 obtained from the SUBARU dataset, zoomed at 
the level of the HB.
The main features of this CMD have been presented in \citep{ripepi07} and
will be further discussed in Federici et al. (2008, in preparation).
Here we focus on the discovery of many new variable stars in the cluster, and 
of several new first-overtone (RRc) RR Lyrae stars, in particular. 
\section{Some results}
Periods and classification in types for the new candidate variables 
(identified with independent techniques, details in Di Criscienzo et al. 2007, in preparation)
were derived using GRaTiS (Graphycal Analizer of Time Series), 
a custom software developed at the Bologna Observatory (see Clementini et al. 2000). 
Reliable periods were obtained for 100 stars: 75 RR Lyrae, 1 Population II Cepheid, 
11 SX Phoenicis, 2 $\delta$ Scuti, 3 binaries, 5 long period variables near the 
Tip of the Red Giant Branch, plus 3 more candidate variables not classified yet. 
The newly discovered RR Lyrae variables include 11 fundamental-mode (RRab) 
and 28 first-overtone (RRc) pulsators (see Fig.1). Examples of the $V$ and $B$ light curves 
of RRab and RRc variables are shown in Fig. 2.
With these new discoveries, the ratio of the number of RRc stars over the number 
of RRc+RRab variables changes from 0.28 (based on the census by Pinto \& Rosino, 1977) 
to 0.49, in much better agreement with the expectations for a metal-poor cluster, and with 
the classification of NGC 2419 as an Oosterhoff type II (OoII) cluster. A classification confirmed 
and strengthen by the average period of the RRab stars and the period-amplitude distribution 
derived in our study (see Ripepi et al., 2007; Di Criscienzo et al. 2007).
 The pure OoII nature of NGC 2419 seems to disfavor the possibility of an extragalactic origin 
 for the cluster, since  field and cluster RR Lyrae stars in extragalactic systems generally
 have properties  intermediate between the two Oosterhoff types 
 (e.g. Catelan 2004). However, since five of the Large Magellanic Cloud GCs 
 also have OoII type, and also Ursa Minor and Bootes I, among the dwarf spheroidal 
 galaxies are pure OoII systems, an extragalactic origin of NGC 2419 may not be totally 
 ruled out. On the other hand, as noted by Ripepi et al. (2007), the apparent lack of multiple stellar 
populations and of metallicity spreads does not corroborate the hypothesis that NGC 2419 
might be the core of a defunct galaxy. From our data, this cluster appears indeed 
a "normal", low metallicity Galactic GC. The only exceptional feature is the presence of 
the HB "blue-hook'', a feature that, up to now, has been detected only in very few Galactic 
GCs, all showing signs of multiple stellar populations, such as NGC 2808 (Piotto et
al., 2007) and/or an extragalactic origin. Interestingly, the occurrence of an extended HB
seems to 
correlate with the mass of the cluster (Recio-Blanco et al. 2006). Indeed, NGC 2419 is 
one of the 
most massive Galactic GCs.
Our next step will be to investigate this HB "anomaly", and to study in the detail 
the CMD to confirm or discard the "normal" nature of NGC 2419 as a Galactic GC, suggested by 
the properties of the cluster RR Lyrae stars.
\begin{figure}[]
\resizebox{\hsize}{!}{\includegraphics[clip=true]{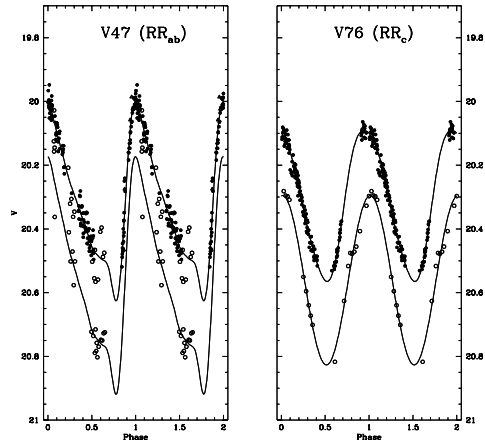}}
\caption{
\footnotesize
$V$ (upper curve) and $B$ (lower curve) light curves of two new RR Lyrae stars
discovered in NGC 2419. Lines are models obtained by properly scaling down the star's $V$ 
light curve.
}
\label{li_vhel}
\end{figure}

\bibliographystyle{aa}

\end{document}